\documentclass{article}
\usepackage{amssymb}
\usepackage{amsmath}
\usepackage{amsthm}

\begin{document}

\title{The self-concordant perceptron is efficient on a sub-family feasibility instances.}

\author{Adrien CHAN-HON-TONG \\ ONERA Universit\'e Paris Saclay, 91140, France}

\maketitle

\subsection*{Abstract}

Strict linear feasibility or linear separation is usually tackled using efficient approximation/stochastic algorithms (that may even run in sub-linear times in expectation). However, today state of the art for solving exactly/deterministically such instances is to cast them as a linear programming instances. Inversely, this paper introduces a self-concordant perceptron algorithm which tackles directly strict linear feasibility with interior point paradigm. This algorithm has the same worse times complexity than state of the art linear programming algorithms but it complexity can be characterized more precisely eventually proving that it binary complexity is low on a sub-family of linear feasibility.

\section{Introduction}
Linear separation is the problem of solving a system of linear inequality $Ax\geq \mathbf{1}$ where $A\in \mathbb{R}^{M\times N}$ is a given matrix and $x$ is a $N$ vector to find (assuming some exist).
Trivially, this problem is equivalent to strict linear feasibility which aims to solve $Ax> \mathbf{0}$ as any solution of the strict linear feasibility can be up-scaled to become a solution of linear separation.

This problem can be found in many domains notably machine learning where it usually comes with additional regularization cost on $x$ leading to linear support vector machine (SVM) for binary classification: given $x_1, ..., x_M \in \mathbb{R}^N$ and $y_1, ..., y_M\in\{-1,1\}$, the hyperplane that maximizes the margin between positive and negative samples (assumed linearly separable) is given by \[\underset{w \in \mathbb{R}^N,b\in \mathbb{R}}{\min} w^Tw \ \ \ \mathrm{sb:} \ \forall m \in \{1,...,M\},\ y_m(w^Tx_m+b)\geq 1\]
Trivially, if one drops the $w^Tw$ costs, the problem becomes to find $w,b$ such that $\forall m \in \{1,...,M\},\ y_m(w^Tx_m+b) = y_m((w::b)^T(x_m::1)) = (w::b)^T(y_m\times x_m::y_m))\geq 1$ which is a linear separation problem with $A$ the matrix $M\times(N+1)$ whose row $m$ is $y_m\times x_m::y_m$ where $::$ stands for the concatenation.

This problem is usually approximated very efficiently in expectation. For example \cite{shalev2007pegasos} proves that one can get a given approximation value with a number of iteration independent of $M$. Also, \cite{joachims2006training} offers a linear time approximation algorithm for linear SVM.

Now, for exact/deterministic solution, methods dedicated to linear feasibility like \cite{dunagan2008simple,pena2016deterministic} does not match the state of the art of linear programming.
For example, \cite{pena2016deterministic} may require $O(NM^2\sqrt{M} \log(\frac{1}{\rho}))$ operations with $\rho$ being the maximal margin related to the instance i.e. $\rho = \underset{x\in \mathbb{R}^N, x^Tx\leq 1}{\max} \ \underset{m\in \{1,...,M\}}{\min} \ A_mx$ (in other words $\frac{1}{\rho}$ is the norm of the SVM related to $A$). Yet, if the matrix $A$ has total binary size $L$, this correspond to $O(NM^2\sqrt{M} L)$ in worse case, while state of the art of linear programming is currently $N^\gamma \sqrt{M} L$ operations (where $\gamma$ is the coefficient of matrix multiplication currently around 2.38) with \cite{renegar1988polynomial,nesterov1994interior} and even $N^{\gamma-\frac{1}{2}}\sqrt{M} L$ with \cite{van2020deterministic} a deterministic version of \cite{cohen2019solving} which offers a data structure allowing to compress the time complexity of the sequence of matrix inversions (same number of iterations but with amortized lighter matrix inversions).

Thus, today, the best theoretical way to deal (exactly and deterministically) with a strict linear feasibility instance $Ax> \mathbf{0}$ is to cast it into a linear program like: \[\underset{x\in \mathbb{R}^N,t\in \mathbb{R}}{\min} \ t \ \ \ \ \mathrm{sb:} \ Ax+t\mathbf{1}\geq \mathbf{0}, \ \ -\mathbf{1}\leq x\leq \mathbf{1} \]
This last problem is feasible with a trivial starting point $x=0,t=1$ and it is bounded with $t$  coarsely able to become $-\rho$ (it would be exactly $-\rho$ if the constraint was $x^Tx\leq 1$ instead of $-\mathbf{1}\leq x\leq \mathbf{1}$). 

Then, this last problem can be solved classically by interior point method i.e. by considering \[
G_C(x,t) = C\times t - \underset{m\in\{1,...,M\}}{\sum} \log(A_mx+t) \]\[\ \ \ \ \ \ \ \ \ \ \ \ \ \ \ \ \ \ - \underset{m\in\{1,...,N\}}{\sum} \log(1+x_n)+\log(1-x_n)
\]
Optimizing $x,t$ to reach $\underset{x,t}{\min} \ G_C(x,t)$ then increasing $C$ as when $C$ is large enough on can guarantee $t<0$.

Even more, if $L$ is the total binary size of $A$, one can prove that the calculus of this algorithm never require more than $O(L)$ bit for each variable, which is optimal in worse case.

\section{Motivation}
Despite binary property of state of the art linear programming algorithms are optimal in worse cases, no works have focused on characterizing the instances where binary size will be significantly lower than $L$. Indeed, even when convergence is very fast, binary size required at each step may be high as this algorithm (i.e. minimization of $G_C$) structurally increases $C$ precisely in order to allow the current point to go closer and closer to the boundary of the polyhedron $\{x,\ Ax+t\geq \mathbf{1}, \ -\mathbf{1}\leq x\leq \mathbf{1}\}$. This makes it hard to bound precisely the binary dynamic of the variables through the resolution of the problem. 
In other words, the classical algorithms work by making the log converging toward a perfect barrier that would be 0 on strict positive values and infinity on 0 and negative ones, eventually leading some term to go very close to 0 even when encapsulated into a $\log$ and thus requiring careful rounding strategy.

Inversely, this paper offers an algorithm with same worse time complexity but  whose binary times complexity can be precisely characterized: let $\mu = \underset{i,j}{\sum} |A_iA_j^T| $ and $\rho$ the margin (i.e.  $\underset{x\in \mathbb{R}^N, x^Tx\leq 1}{\max} \ \underset{m\in \{1,...,M\}}{\min} \ A_mx$), then, the offered algorithm requires $O(\sqrt{M}\log(\frac{\sqrt{M}\mu}{\rho}))$ Newton steps with each variable having almost always a denominator bounded by $O(M)$ and a numerator bounded by $O(\log(\frac{\sqrt{M}\mu}{\rho}))$. In particular, for almost square ($N\approx M$) matrix densely filled with large number $O(\log(\mu)) \approx \frac{L}{M^2}$ rather than $L$. If in addition the margin is high such that $\log(\frac{1}{\rho}) \approx \frac{L}{M}$, then this algorithm is theoretically better by a factor $\frac{1}{M}$ for the binary size required during the algorithm than the conversion to a linear program where operations have not been proven possible with less than $L$ digits (as pointed table 1).
\begin{table}[ht]
    \centering
    \begin{footnotesize}
    \begin{tabular}{c||c|c}
         & classical way & offered algorithm \\\hline\hline
        worse number of iteration & $\sqrt{M}L$ & $\sqrt{M}L$ \\\hline
        precise number of iteration & unclear & $O(\sqrt{M}\log(\frac{\sqrt{M}\mu}{\rho}))$ \\\hline\hline
        worse number of bits/variable & $L$ & $L$ \\\hline
        precise number of bits/variable & unclear & $O(\log(\frac{\sqrt{M}\mu}{\rho}))$ \\\hline\hline
        binary complexity on almost square & unclear & know to be\\
        dense instance with high margin & ($\leq O(M^\gamma\sqrt{M}L^2)$) & $O(M^{\gamma-2}\sqrt{M}L^2)$  \\\hline\hline
        behavior of log influence & decreases smoothly & increase smoothly \\\hline
        known worse cases & yes \cite{allamigeon2022no} & not yet
    \end{tabular}
    \end{footnotesize}
    \caption{Complexity for solving a linear feasibility query $Ax>\mathbf{0}$ with $A\in \mathbb{R}^{M\times N}$ with total binary size $L$ by either solving the problem with state of the art linear program algorithms or with the self-concordant perceptron. }
\end{table}

This algorithm has also two other interests:
\begin{itemize}
    \item The algorithm works by making the log terms more and more important rather then less and less important, this way it may not struggle on the same hard instances than classical path following algorithm which has currently be proven not strongly polynomial \cite{allamigeon2022no} on some hard instances. This may also be connected to the idea of converging to the center of the polyhedron rather than a vertex \cite{inayatullah2019new}.
    \item The proof is somehow simpler than the other of central-path based methods (no need to introduced neither the complementary condition nor the distance to the central path).
\end{itemize}

\textit{Let stress that the self-concordant perceptron described in this paper is a significantly improvement of the version from \cite{chan2022new} where complexity was $O(ML)$ and not $O(\sqrt{M}L)$ thank to the adaptation of the path following process to this algorithm.}

\section{The self-concordant perceptron}

\subsection{Well-known theorems}
This paper relies on the following well-known theorems.

\subsubsection{Determinant bound:}
If $A$ is a square matrix with total binary size $L$, then, $\log(Det(A))\leq L$. As a result, if $A$ is a matrix with total binary size $L$ (not necessarily square) and $\widetilde{A}$ is a square sub-matrix of $A$ with possibly $O(1)$ new rows and cols with $O(1)$ values, then, $\log(Det(\widetilde{A}))\leq L$. Combined with Cramer rules, it also implies that it $\chi$ verifies $\widetilde{A}x = \mathbf{1}$ (or any vector with $O(1)$ values), then, $\log(\chi^T\chi)\leq L$.
This result is used to establish time complexity of linear programming algorithm since at least \cite{karmarkar1984new}.
In particular, $\log\left(\underset{x, \ Ax \geq \mathbf{1}}{\min} x^Tx \right) \leq O(L)$.

\subsubsection{Self concordance theory:}
If $\Psi(x)$ is a self-concordant function (mainly any sum of quadratic, linear, constant and $-\log$ term), with a minimum $\Psi^*$, then, the Newton descent starting from $x_{start}$ allows to compute $x_\epsilon$ such as $\Psi(x_\epsilon)-\Psi^*\leq \varepsilon$ in $\widetilde{O}(\Psi(x_{start})-\Psi^*+\log\log(\frac{1}{\varepsilon}))$ damped Newton steps. Each step is: \begin{itemize}
    \item $\lambda_\Psi(x) \leftarrow \sqrt{(\nabla_x\Psi)^T(\nabla^2_x\Psi)^{-1}(\nabla_x\Psi)}$
    \item $x \leftarrow x - \frac{1}{1+\lambda_\Psi(x)}(\nabla^2_x\Psi)^{-1}(\nabla_x\Psi)$
\end{itemize}
Importantly, $(x-x^*)^T(\nabla^2_x\Psi)^{-1}(x-x^*)\leq \frac{\varepsilon}{1-\varepsilon}$ when when $\Psi(x_\epsilon)-\Psi^*\leq \varepsilon$ (for $\varepsilon$ below 1).
The proof of this central theorem can be found in \cite{nemirovski2004interior,boyd2004convex}.

\subsection{The self concordant Perceptron}

The self concordant Perceptron is described as a pseudo code in table 2.
\begin{table}[ht]
\texttt{def \ self-concordant-perceptron}$(A)$ :
\begin{enumerate}
    \item $\varepsilon \leftarrow \frac{1}{2^{O(L)}}$, $\delta \leftarrow 1$, $v\leftarrow \mathbf{1}$, $\mu \leftarrow \underset{i,j}{\sum} |A_iA^T_j|$
    \item while $\neg (AA^Tv>\mathbf{0})$
    \begin{enumerate}
        \item symbolically $\ F_\delta(u) = \delta \times \mathbf{1}^Tu + \frac{u^TAA^Tu}{\mu} - \underset{m\in\{1,...,M\}}{\sum} \log(u_m)$
        \item $v\leftarrow$ rounded Newton descent from $v$ on $F_\delta(u)$ with $\varepsilon$ precision
        \item $\delta\leftarrow \delta \times (1-\frac{1}{\sqrt{M}})$
    \end{enumerate}
\end{enumerate}

    \caption{Pseudo code of the self concordant perceptron.}
    
\end{table}

More mathematically, given $A\in \mathbb{Z}^{M\times N}$ a feasibility instance with $\mu = \underset{i,j}{\sum}|A_iA_j^T|$, the algorithm is mainly a Newton descent on the function $\forall v,\ F_\delta(v) = \delta \times \mathbf{1}^Tv + \frac{v^TAA^Tv}{\mu} - \underset{m\in\{1,...,M\}}{\sum} \log(v_m)$ to reach $\min_v F_\delta(v)$ then decaying $\delta$ to move into the central-path related to $F_\delta$.

To prove the claim presented in table 1, this paper will proceed as follow:
\begin{itemize}
    \item first, it will be proven that the central path converge toward a solution,
    \item then, the complexity of the algorithm will be established from mathematical point of view (without rounding consideration),
    \item finally, it will be proven that the rounding does not harm the process.
\end{itemize}

\subsection{Convergence of the central path}
\subsubsection{Lemma on the lower bound of the quadratic term} Let  $A\in \mathbb{Z}^{M\times N}$ with the assumption  that there is a solution to $Ax\geq \mathbf{1}$ and let $x$ the SVM related to $A$ i.e. $x = \underset{\chi \in \mathbb{R}^N}{\arg\min}\ \chi^T\chi \ \ \ \ \mathrm{sb:} Ax\geq \mathbf{1}$.

then, $\forall v\geq \mathbf{0}$, then, $\frac{||v||^2_2}{||x||^2_2} \leq ||A^Tv||^2_2$, and, thus  $||A^Tv||_2^2\leq M\mu \Rightarrow v\leq \frac{\sqrt{M} \mu}{\rho} \times \mathbf{1}$ where $\rho = \frac{1}{\sqrt{x^Tx}}$ is the margin related to $A$.


\paragraph{proof}
Cauchy inequality applied to $x^T(A^Tv)$ gives: $x^T(A^Tv)\leq ||x||_2\times ||A^Tv||_2$.

But, $x^T(A^Tv) = (Ax)^Tv\geq \mathbf{1}^Tv$ as $v\geq \mathbf{0}$ and $Ax\geq \mathbf{1}$.
Thus, $\mathbf{1}^Tv \leq ||x||_2\times ||A^Tv||_2$.

As each side is positive, one could take the square (and push $||x||_2$ to the left), this gives $\frac{(\mathbf{1}^Tv)^2}{||x||^2_2} \leq ||A^Tv||^2_2$.
Yet, as $v\geq \mathbf{0}$, $v^Tv\leq (\mathbf{1}^Tv)^2$.

This inequality directly implies the second by injecting of $||A^Tv||_2^2\leq M\mu$.


\subsubsection{Optimality conditions}
$\forall \delta>0, v>\mathbf{0}, v^TAA^Tv\geq 0$ thus $ F_\delta(v) = \delta \times \mathbf{1}^Tv + \frac{v^TAA^Tv}{\mu} - \underset{m\in\{1,...,M\}}{\sum} \log(v_m)\geq \underset{m\in\{1,...,M\}}{\sum} \delta v_m - \log(v_m)$. Thus, $F_\delta(v)$ is trivially lower bounded. As $F_\delta$ is convex, it has a minimum $w(\delta) = \underset{v\in \mathbb{R}^M}{\arg\min } \ F_\delta(v)$.

Now, from null gradient condition, it comes that 
\[\forall \delta>0, m\in \{1,...,M\}, \delta + 2\frac{A_mA^Tw(\delta)}{\mu} -\frac{1}{w_m(\delta)} =0 \]

In particular, $\forall \delta>0, m\in \{1,...,M\}, \delta \times w_m(\delta) + 2w_m(\delta) \times \frac{A_mA^Tw(\delta)}{\mu} =1 $. By summing over $m$, $\forall \delta>0, \delta \times \mathbf{1}^Tw(\delta) + 2\frac{w(\delta)^TA_mA^Tw(\delta)}{\mu} =M $ thus, $2w(\delta)^TAA^Tw(\delta) < M \mu $ as $w>\mathbf{0}$ implies $\mathbf{1}^Tw(\delta)>0$.

\subsubsection{Solution}
Combining lemma 3.3.1 and 3.3.2, it comes that $\forall \delta>0, m\in \{1,...,M\}$
\[w_m(\delta)\leq \frac{\sqrt{M}\mu}{\rho}, \mathrm{and},\ \delta + \frac{2A_mA^Tw(\delta)}{\mu} -\frac{1}{w_m(\delta)} =0 \]

In particular, for $\forall \delta>0$ \[\delta < \frac{\rho}{\sqrt{M}\mu} \Rightarrow \ \frac{2A_mA^Tw(\delta)}{\mu} = \frac{1}{w_m(\delta)} - \delta > 0 \]  
\[\forall 0<\delta < \frac{\rho}{\sqrt{M}\mu}, \ AA^Tw(\delta)>\mathbf{0}\]

Importantly, the existence of $w(\delta)$ is trivial proven for any $\delta>0$, but the fact that $w$ converges toward a solution is \textbf{only} true when there exists a solution to the linear feasibility instance (i.e.  $w(0)$ exists) as it requires the inequality 3.3.1: for large $\delta$ the linear term forbids $w$ to be too large but allows for not being a solution, while for small $\delta$, the quadratic term takes over the role of preventing $w$ to be large yet forcing it to converge toward a solution as $\delta$ is then negligible.


\subsubsection{Number of decaying steps}
As the algorithm stops at least when $\delta\leq \frac{\rho}{\sqrt{M}\mu}$, and that each step decay $\delta$ by a factor $(1-\frac{1}{\sqrt{M}})$, then, the number of steps verifies $(1-\frac{1}{\sqrt{M}})^K\leq  \frac{\rho}{\sqrt{M}\mu}$ i.e. $k \leq \sqrt{M}\log(\frac{\sqrt{M}\mu}{\rho})$.

However, at this point step $c$ may require many Newton steps.
But, this paper proves that in average (deterministically) it does not.

\subsection{Complexity}
\subsubsection{Initialization} 
By definition, $F_1(v) = \mathbf{1}^Tv + \frac{v^TAA^Tv}{\mu}-\underset{m\in \{1,...,M\}}{\sum} \log(v_m) \geq \underset{m\in \{1,...,M\}}{\sum} v_m - \log(v_m) \geq M$ as the function $\psi(u) = u-\log(u)$ verifies $\psi'(u) = 1-\frac{1}{u}$ eventually $\phi(u)\geq \phi(1)=1$.

Independently, $F_1(\mathbf{1}) \leq M+1$.
Thus, it takes only $O(1 + \log\log(\frac{1}{\varepsilon}))$ to get a $\varepsilon$-approximation of $w(1)$ from $\mathbf{1}$.

\subsubsection{Centering cost}
Let first remark that optimality condition 3.3.2 not only implies that $\frac{w(\delta)AA^Tw(\delta)}{\mu}\leq M$ as $\delta \mathbf{1}^Tw(\delta)>0$ but, also that $\delta \mathbf{1}^Tw(\delta)<M$ as $w(\delta)AA^Tw(\delta)>0$.  Then

\[F_{(1-\frac{1}{\sqrt{M}})\delta}(w(\delta)) = F(w(\delta),\delta)+((1-\frac{1}{\sqrt{M}})\delta - \delta)\mathbf{1}^Tw(\delta)\]
\[= F(w(\delta),\delta)-\frac{\delta\mathbf{1}^Tw(\delta)}{\sqrt{M}} \]

But, $F_{\delta}( w((1-\frac{1}{\sqrt{M}})\delta)= F_{(1-\frac{1}{\sqrt{M}})\delta}( w((1-\frac{1}{\sqrt{M}})\delta)$ 

$ + (\delta - (1-\frac{1}{\sqrt{M}})\delta)\mathbf{1}^T  w((1-\frac{1}{\sqrt{M}})\delta)$

$= F_{(1-\frac{1}{\sqrt{M}})\delta}( w((1-\frac{1}{\sqrt{M}})\delta)$

$ + (\frac{1}{1-\frac{1}{\sqrt{M}}} - 1) \times (1-\frac{1}{\sqrt{M}})\delta  \mathbf{1}^T  w((1-\frac{1}{\sqrt{M}})\delta)$

$= F_{(1-\frac{1}{\sqrt{M}})\delta}( w((1-\frac{1}{\sqrt{M}})\delta)$

$ + (\frac{\sqrt{M}}{\sqrt{M}-1} - \frac{\sqrt{M}-1}{\sqrt{M}-1}) \times (1-\frac{1}{\sqrt{M}})\delta  \mathbf{1}^T  w((1-\frac{1}{\sqrt{M}})\delta)$

$= F_{(1-\frac{1}{\sqrt{M}})\delta}( w((1-\frac{1}{\sqrt{M}})\delta)$
$ + \frac{(1-\frac{1}{\sqrt{M}})\delta  \mathbf{1}^T  w((1-\frac{1}{\sqrt{M}})\delta)}{\sqrt{M}-1} $

By combining the two 
\[F_{(1-\frac{1}{\sqrt{M}})\delta}(w(\delta)) =  F_{(1-\frac{1}{\sqrt{M}})\delta}( w((1-\frac{1}{\sqrt{M}})\delta) \]\[ -\frac{\delta\mathbf{1}^Tw(\delta)}{\sqrt{M}} + \frac{(1-\frac{1}{\sqrt{M}})\delta  \mathbf{1}^T  w((1-\frac{1}{\sqrt{M}})\delta)}{\sqrt{M}-1} \]

Thus, the cost of an individual step \textit{2.b} of algorithm table 1 is \[O\left(\frac{(1-\frac{1}{\sqrt{M}})\delta  \mathbf{1}^T  w((1-\frac{1}{\sqrt{M}})\delta)}{\sqrt{M}-1} -\frac{\delta\mathbf{1}^Tw(\delta)}{\sqrt{M}} + \log\log(\frac{1}{\varepsilon})\right)\] Newton steps, and, this can be as large as $\sqrt{M}$ if $ (1-\frac{1}{\sqrt{M}})\delta  \mathbf{1}^T  w((1-\frac{1}{\sqrt{M}})\delta)=O(M)$ and $\delta\mathbf{1}^Tw(\delta)=O(1)$.

But, let first remark that if $(1-\frac{1}{\sqrt{M}})\delta  \mathbf{1}^T  w((1-\frac{1}{\sqrt{M}})\delta)\leq \delta\mathbf{1}^Tw(\delta)$ (i.e. the quantity $\delta\mathbf{1}^Tw(\delta)$ decreases after $\delta$ decay), then, such step only costs at most $O(1 + \log\log(\frac{1}{\varepsilon}))$ as $M(\frac{1}{\sqrt{M}-1} - \frac{1}{\sqrt{M}}) = \frac{M}{M-\sqrt{M}} \underset{M\rightarrow\infty}{\rightarrow}1$. 

And, on the other hand, $F_{(1-\frac{1}{\sqrt{M}})\delta}(w(\delta)) \leq  F_{(1-\frac{1}{\sqrt{M}})\delta}( w((1-\frac{1}{\sqrt{M}})\delta)$.
Thus, \[\frac{(1-\frac{1}{\sqrt{M}})\delta  \mathbf{1}^T  w((1-\frac{1}{\sqrt{M}})\delta)}{\sqrt{M}-1} -\frac{\delta\mathbf{1}^Tw(\delta)}{\sqrt{M}} \geq 0\]
\[(1-\frac{1}{\sqrt{M}})\delta  \mathbf{1}^T  w((1-\frac{1}{\sqrt{M}})\delta) \geq (1-\frac{1}{\sqrt{M}}) \times \delta\mathbf{1}^Tw(\delta) \geq \delta\mathbf{1}^Tw(\delta) - \sqrt{M}\]

Thus, a $\sqrt{M}$-cost step (corresponding to the worse case $\delta\mathbf{1}^Tw(\delta)=O(1)$ while $(1-\frac{1}{\sqrt{M}})\delta  \mathbf{1}^T  w((1-\frac{1}{\sqrt{M}})\delta)=O(M)$) can only happen after $\sqrt{M}$ decreasing steps from an hypothetical situation where $\delta\mathbf{1}^Tw(\delta)=O(M)$.

So if one consider not only a single step but the set of all steps where large increase of $\delta \mathbf{1}^Tw(\delta)$ between $\delta$ and $(1-\frac{1}{\sqrt{M}})\delta$ should necessarily be preceded by large number of steps with decreasing $\delta \mathbf{1}^Tw(\delta)$, then the amortized cost of a step is $O(1 + \log\log(\frac{1}{\varepsilon}))$.

In other words, for any given step where $\delta \mathbf{1}^Tw(\delta)$ is decreasing, the algorithm should spend twice $O(1)$ Newton steps.
One $O(1)$ Newton steps to go from $w(\delta)$ to $w((1-\frac{1}{\sqrt{M}})\delta)$, and, one $O(1)$ in the future to return from $(1-\frac{1}{\sqrt{M}})\delta \mathbf{1}^Tw((1-\frac{1}{\sqrt{M}})\delta)$ level to $\delta \mathbf{1}^Tw(\delta)$. 
Yet, it is still $O(1)$ when amortizing.

Importantly, it is not a $O(1)$ on expectation (which would be related to a randomized algorithm), it is $O(1)$ with one part that would be done after.

\subsubsection{Mathematical complexity}
As step \textit{2.b} costs $O(1+\log\log(\varepsilon))$ in average, then, the total complexity is $O\left( M^{\gamma} \sqrt{M}\log(\frac{\sqrt{M}\mu}{\rho})(1+\log\log(\frac{1}{\varepsilon})) \right)$ which is $O(M^{\gamma}\sqrt{M}L)$ in worse case, but which is better characterized.

In particular, on instance where $\log(\frac{\mu}{\rho})\ll L$, then, this algorithm is interesting.

Now, the classical way may also benefit from the fact that $\log(\frac{\mu}{\rho})\ll L$.
Yet, this algorithm has an other interesting property: $v\leq \log(\frac{\mu}{\rho})$ and all log terms are large. Yet, this last property seems structurally hard for the classical way where log terms tend to $0$.

Thus, complexity of this algorithm is equivalent to the one of the classical way, but with a better characterization which may however be shared with the classical way. But, the binary property of this algorithm may be unique in those last \textit{easy} instances.

\subsection{Effect of rounding}

\subsubsection{variable rounding}
The interest of this algorithm is that ceiling $v$ during Newton step can be almost trivially.
Indeed, $\underset{m}{\sum} \log(v_m+\zeta_m)>\underset{m}{\sum} \log(v_m)$ if $\zeta\geq \mathbf{0}$
Thus, to design a ceiling process, one only has to consider $\delta\mathbf{1}^T(v+\zeta) + \frac{(v+\zeta)^TAA^T(v+\zeta)}{\mu}$ with the additional knowledge that $\frac{v^TAA^Tv}{\mu}\leq M$.

Thus $\delta\mathbf{1}^T(v+\zeta) + \frac{(v+\zeta)^TAA^T(v+\zeta)}{\mu} - \delta\mathbf{1}^Tv + \frac{v^TAA^Tv}{\mu} = \delta \mathbf{1}^T\zeta + \frac{\zeta^TAA^T\zeta}{\mu} + 2\frac{\zeta^TAA^Tv}{\mu} \leq M ||\zeta||_\infty +  ||\zeta||_\infty^2 + 2M \frac{||A\zeta||}{\sqrt{\mu}} \leq 3M ||\zeta||_\infty + ||\zeta||_\infty^2$

Thus, a simple periodic ceiling with denominator $O(M)$ is sufficient to maintain a $O(1)$ errors which can be compensated with $O(1)$ Newton steps.

\subsubsection{weight rounding}
In the same way $(\delta + \zeta)\mathbf{1}^Tv$ can be harmless if $\zeta\leq \frac{\delta}{M}$ as in this case, $(\delta + \zeta)\mathbf{1}^Tv - \delta\mathbf{1}^Tv \leq \frac{\delta\mathbf{1}^Tv}{M} \leq 1$.

\bibliographystyle{plain}
\bibliography{ref}

\begin{thebibliography}{10}

\bibitem{allamigeon2022no}
Xavier Allamigeon, St{\'e}phane Gaubert, and Nicolas Vandame.
\newblock No self-concordant barrier interior point method is strongly
  polynomial.
\newblock In {\em Proceedings of the 54th Annual ACM SIGACT Symposium on Theory
  of Computing}, pages 515--528, 2022.

\bibitem{boyd2004convex}
Stephen~P Boyd and Lieven Vandenberghe.
\newblock {\em Convex optimization}.
\newblock Cambridge university press, 2004.

\bibitem{chan2022new}
Adrien Chan-Hon-Tong.
\newblock A new algorithm for linear programming in critical systems.
\newblock {\em SN Computer Science}, 4(1):76, 2022.

\bibitem{cohen2019solving}
Michael~B Cohen, Yin~Tat Lee, and Zhao Song.
\newblock Solving linear programs in the current matrix multiplication time.
\newblock In {\em Proceedings of the 51st annual ACM SIGACT symposium on theory
  of computing}, 2019.

\bibitem{dunagan2008simple}
John Dunagan and Santosh Vempala.
\newblock A simple polynomial-time rescaling algorithm for solving linear
  programs.
\newblock {\em Mathematical Programming}, 114(1):101--114, 2008.

\bibitem{inayatullah2019new}
Syed Inayatullah, Maria Aman, Asma Rani, Hina Zaheer, and Tanveer~Ahmed
  Siddiqi.
\newblock A new technique for determining approximate center of a polytope.
\newblock {\em Advances in Operations Research}, 2019:1--7, 2019.

\bibitem{joachims2006training}
Thorsten Joachims.
\newblock Training linear svms in linear time.
\newblock In {\em Proceedings of the 12th ACM SIGKDD international conference
  on Knowledge discovery and data mining}, pages 217--226, 2006.

\bibitem{karmarkar1984new}
Narendra Karmarkar.
\newblock A new polynomial-time algorithm for linear programming.
\newblock In {\em Proceedings of the sixteenth annual ACM symposium on Theory
  of computing}, 1984.

\bibitem{nemirovski2004interior}
Arkadi Nemirovski.
\newblock Interior point polynomial time methods in convex programming.
\newblock {\em Lecture notes}, 42(16):3215--3224, 2004.

\bibitem{nesterov1994interior}
Yurii Nesterov and Arkadii Nemirovskii.
\newblock {\em Interior-point polynomial algorithms in convex programming}.
\newblock Siam, 1994.

\bibitem{pena2016deterministic}
Javier Pe{\~n}a and Negar Soheili.
\newblock A deterministic rescaled perceptron algorithm.
\newblock {\em Mathematical Programming}, 155(1-2):497--510, 2016.

\bibitem{renegar1988polynomial}
James Renegar.
\newblock A polynomial-time algorithm, based on newton's method, for linear
  programming.
\newblock {\em Mathematical programming}, 40(1):59--93, 1988.

\bibitem{shalev2007pegasos}
Shai Shalev-Shwartz, Yoram Singer, and Nathan Srebro.
\newblock Pegasos: Primal estimated sub-gradient solver for svm.
\newblock In {\em Proceedings of the 24th international conference on Machine
  learning}, pages 807--814, 2007.

\bibitem{van2020deterministic}
Jan van~den Brand.
\newblock A deterministic linear program solver in current matrix
  multiplication time.
\newblock In {\em Proceedings of the Fourteenth Annual ACM-SIAM Symposium on
  Discrete Algorithms}, pages 259--278. SIAM, 2020.

\end{thebibliography}

\section{APPENDIX}
This appendix is an other interest of the previous framework: the possibility to check if a relevant path following from any point.

This appendix is intented to be somehow self contained.

\subsection{Problem}
Given a matrix $A\in \mathbb{Q}^{M\times N}$ with total binary size $L$ and the assumption that there exists $x\in \mathbb{Q}^N$ such that $Ax>\mathbf{0}$, one can look for an algorithm for producing such $x$ (which can alternatively be computed as $A^Tv$ with $v\in \mathbb{Q}^M$ such as $AA^Tv>\mathbf{0}$).

There are many algorithms requiring no more than $O(\sqrt{M}L)$ \textit{steps} (each of them being typically a matrix inversion e.g. a damped Newton step) to converge.

The most common of such algorithm is path following: moving from successive minimums of derived problems.
However, path following method usually comes with the limitation that exiting the corridor around the central path may eventually broke the algorithm convergence and/or complexity.

This appendix introduces a \textit{common} method but with the ability to start a path following from any point, and, with the relevancy to do so well caracterized.

\subsection{Background results}
The following results will be used without being proved as part of the common knowledge.

\subsubsection{Self concordance theory}
A self concordance function $\mathcal{F}$ with a minimum $x^*$ and a minimal value $\mathcal{F}^*=\mathcal{F}(x^*)$ can be minimized (i.e. producing $x$ with $\mathcal{F}(x)-\mathcal{F}^*\leq \varepsilon$) from any point $x_0$ in less than $O(\mathcal{F}(x_0)-\mathcal{F}^*+\log\log(\frac{1}{\varepsilon})$ damped Newton steps.

And, any positive quadratic + linear + sum of $-\log$ terms is a self concordant function.

\subsubsection{Cramer related bounds}
If there exists $x$, $Ax>\mathbf{0}$, then there exists $x, Ax\geq \mathbf{1}$ (this is trivial by dividing by the strictly positive $\min_m A_mx$), and, one such $x$ can be written like $(A_\mathcal{S})^{-1}\mathbf{1}$. So each component of $x$ can be written using Cramer formula.

A fortiori, this $x$ verifies that $||x||$ is smaller than the maximal sub determinant of $||A||$.

Finally, the log of this maximal sub determinant is lower than $O(L)$ where $L$ is the binary size of $A$.

\subsubsection{Cauchy related bounds}
If there exists $x$, $Ax\geq 1$, then $\forall v\geq \mathbf{0}$, $(v^TA)x= v^T(Ax) \geq v^T\mathbf{1}$, and independently, $(v^TA)x\leq \sqrt{v^TAA^Tv} \sqrt{x^Tx}$ by Cauchy.

So, $\forall v\geq \mathbf{0},\ 0\leq \frac{v^Tv}{x^Tx}\leq \frac{(1^Tv)^2}{x^Tx}\leq v^TAA^Tv$.

In particular, it means that $F(v) = \frac{v^TAA^Tv}{2}-\sum_{\{m \in \{1,...,M\}\}} \log(v_m)\geq \frac{v^Tv}{2x^Tx}-\sum_{\{m \in \{1,...,M\}\}} \log(v_m)\geq M \times \min_{\{u\in \mathbb{R}\}} \left(\frac{u^2}{2x^Tx} - \log(u)\right)\geq -O(M\times \log(x^Tx)) = -O(ML)$.
Let point out that it also implies that if $v^TAA^Tv\leq M$ then, $v\leq \sqrt{Mx^Tx} \mathbf{1}$.

As independently,  there is a good trivial point $F(\sqrt{\frac{M}{\mathbf{1}^TAA^T\mathbf{1}}}\mathbf{1})\leq O(M+M\log(\mathbf{1}^TAA^T\mathbf{1}))=O(M\log(||A||)) = O(ML)$, it means that the self concordant function $F$ can be minimized in less than $O(ML)$ damped Newton steps recovering $AA^Tv^*=\frac{1}{v^*}>\mathbf{0}$.

And the question is to known if this can be done in $O(\sqrt{M}L)$ using path following from any point.

\subsection{A generic path following lemma}
Let now introduce a very important lemma which will be useful for the offered algorithm. This lemma generalize path following to a large variety of situation.
\subsubsection{Claim}
 Assume
\begin{enumerate}
    \item $\mu>2, K\geq 2, t(0)>0,\ \rho=1-\frac{1}{\mu}$ (trivially $\rho\geq \frac{1}{2}$), and, let $t(k+1) = \rho t(k)$ (so $t(k) = \rho^k t(0)$),
    \item for any function $\phi,\psi$ from a vector set $E$ to $\mathbb{R}$, let $\varphi(x,t) = \phi(x)+ t \times \psi(x)$,
    \item for any sequences $z : E^\mathbb{N}$ such that \begin{enumerate}
    \item $\varphi(z(n),t(n)) \leq \varphi(z(n+1),t(n))$
    \item $ t(n)\psi(z(n)) \leq \mu^2 $
\end{enumerate}
\end{enumerate} then
    \[\underset{k=1}{\overset{K}{\sum}} \begin{array}{c}
     \varphi(z(k-1),t(k))  \\
     -\varphi(z(k),t(k))
\end{array} \leq \frac{K+\mu-1}{\rho} + (\rho-1)t(0)\psi(z(0)) \]

\subsubsection{Proof}

\begin{center}
\[S =\underset{k=1}{\overset{K}{\sum}} \begin{array}{c}
     \varphi(z(k-1),t(k))  \\
     -\varphi(z(k),t(k))
\end{array}\]
(this is just the definition of $S$)
\[=\underset{k=1}{\overset{K}{\sum}} \begin{array}{c}
     \varphi(z(k-1),t(k))  \\
      -\varphi(z(k-1),t(k-1))+\varphi(z(k-1),t(k-1))\\
     -\varphi(z(k),t(k-1))+\varphi(z(k),t(k-1))\\
     -\varphi(z(k),t(k))
\end{array}\]
(one has just added and subtracted the same quantity so the sum is kept unchanged)
\[=\underset{k=1}{\overset{K}{\sum}} \begin{array}{c}
     \varphi(z(k-1),t(k))  
      -\varphi(z(k-1),t(k-1))\\\varphi(z(k-1),t(k-1))
     -\varphi(z(k),t(k-1))\\+\varphi(z(k),t(k-1))
     -\varphi(z(k),t(k))
\end{array}\]
(simple reorganization)
\[\leq\underset{k=1}{\overset{K}{\sum}} \begin{array}{c}
     \varphi(z(k-1),t(k))  
      -\varphi(z(k-1),t(k-1))\\+\varphi(z(k),t(k-1))
     -\varphi(z(k),t(k))
\end{array}\]
(because $\varphi(z(k-1),t(k-1))\leq
     \varphi(z(k),t(k-1))$ by assumption 3.a)
\[= \underset{k=1}{\overset{K}{\sum}} \begin{array}{c}
     (t(k)-t(k-1))\psi(z(k-1))\\
     + (t(k-1)-t(k))\psi(z(k))
\end{array}\]
(because $\varphi(x,t) = \phi(x)+t\times \psi(x)$, so the $\varphi(x,t)-\varphi(x,t') = (t-t')\psi(x)$)
\[= \underset{k=1}{\overset{K}{\sum}} \begin{array}{c}
     (\rho-1)t(k-1)\psi(z(k-1))\\
     + (\frac{1}{\rho}-1)t(k)\psi(z(k))
\end{array}\]
(by definition of $t$ and $\rho$)
\[= \underset{k=1}{\overset{K}{\sum}} 
     (\rho-1)t(k-1)\psi(z(k-1))+\underset{k=1}{\overset{K}{\sum}} (\frac{1}{\rho}-1)t(k)\psi(z(k))\]
(simple reorganization)
\[= \underset{k=0}{\overset{K-1}{\sum}} 
     (\rho-1)t(k)\psi(z(k))+\underset{k=1}{\overset{K}{\sum}} (\frac{1}{\rho}-1)t(k)\psi(z(k))\]
(variable change in first sum)
\[= \begin{array}{c}
     (\rho-1)t(0)\psi(z(0))
     \\ + \underset{k=1}{\overset{K-1}{\sum}} 
     (\rho-1)t(k)\psi(z(k))
\end{array}+\begin{array}{c}
     \underset{k=1}{\overset{K-1}{\sum}} (\frac{1}{\rho}-1)t(k)\psi(z(k))
     \\ + (\frac{1}{\rho}-1)t(K)\psi(z(K))
\end{array}\]
(simple reorganization)
\[= 
     (\rho-1)t(0)\psi(z(0))
     + \underset{k=1}{\overset{K-1}{\sum}} 
     (\frac{1}{\rho}+\rho-2)t(k)\psi(z(k))
+ (\frac{1}{\rho}-1)t(K)\psi(z(K))
\]
(simple reorganization)
\[= 
     (\rho-1)t(0)\psi(z(0))
     + \underset{k=1}{\overset{K-1}{\sum}} 
     \frac{(1-\rho)^2}{\rho}t(k)\psi(z(k))
+ (\frac{1}{\rho}-1)t(K)\psi(z(K))
\]
(this step is just remarking that $\rho^2-2\rho+1=(1-\rho)^2$, but, it is critical because now the terms in the main sum are positive leading to)
\[\leq (\rho-1)t(0)\psi(z(0))
     + \underset{k=1}{\overset{K-1}{\sum}}\frac{(1-\rho)^2}{\rho}\times \mu^2 + \frac{1-\rho}{\rho}\times \mu^2 \]
(one can finally use the bound $t(k)\times \psi(z(k))\leq \mu^2$ assumption 3.b )
\[= (\rho-1)t(0)\psi(z(0))
     +(K-1)\frac{1}{\rho} + \frac{1}{\rho}\times \mu \]
(because $(1-\rho)\mu = 1$ by definition of $\rho=1-\frac{1}{\mu}$.)
\end{center}

\subsubsection{Important corollary/remark}

For example if $\mu = \sqrt{M}$, $\phi$ and $\psi$ are self concordant with $\varphi(x,t)$ having minimum regarding $x$ for all $t(n)$, and let $z(n) = x^*(t(n))= \arg\min_x \varphi(x,t(n))$ (implying assumption 3.a because $\varphi(z(n),t(n))$ is then minimal), then, the assumption that $\psi(x^*(t(n)))\leq M \Rightarrow $

$\underset{k=1}{\overset{K}{\sum}}\varphi(x^*(t(k-1)),t(k)) -\varphi(x^*(t(k)),t(k)) \leq O(K+M + |t(0)\psi(z^*(0))|)$

which means that following the central path formed by $K$ successive minimums of $\varphi$ for $t=t(0)$ to $t(K)$ can be done in $O(K+M+ |t(0)\psi(z^*(0))|)$ damped Newton steps seeing 1.3.1 result.

\subsection{Measuring the relevancy of any point}

First, let remark that $F(\lambda \times v)$ is optimal for $\lambda=1$ when $v^TAA^Tv = M$ because $F(\lambda \times v)=\frac{\lambda^2}{2}v^TAA^Tv - M\log(\lambda) - \sum_m \log(v_m) \Rightarrow \lambda v^TAA^Tv-\frac{M}{\lambda}=0$ at minimum which is $\lambda = \sqrt{\frac{M}{v^TAA^Tv}}$. So when minimizing $F$, one can always assume the current point verifies $v^TAA^Tv = M$.

Then, $\forall w>\mathbf{0}$ with  $v^TAA^Tv = M$, one can consider
$G_w(v,t) = F(v) + t \times \frac{(v - w)^T(v - w)}{2}$

\subsubsection{Initializing the path}
$\nabla_v G_w(v,t) = \nabla_v F + t (v-w) = AA^Tv-\frac{1}{v}+ t (v-w)$, a fortiori $\nabla_v G_w(w,t) = \nabla_v F(w)$ which does not depend on $t$

But, $\nabla^2_v G_w(v,t) = \nabla^2_v F + t I$ which means the hessian increases with $t$. A fortiori, there is always some $t(0)\geq ||\nabla_v F(w)||^2$ such that $w$ is close to the minimum of $G_w(v,t(0))$.

\subsubsection{Gradient condition}
$\nabla_v G_w(v,t) = \mathbf{0}$ gives $AA^Tv-\frac{1}{v}+ t (v-w)=\mathbf{0}$.
By taking the scalar product with $v-w$, $(v-w)^T (AA^Tv -\frac{1}{v}) + t (v-w)^T(v-w)=0$

$v^TAA^Tv -M - w^TAA^Tv + w^T\frac{1}{v}  + t (v-w)^T(v-w)=0$

$v^TAA^Tv + w^T\frac{1}{v}  + t (v-w)^T(v-w)=M + w^TAA^Tv$

Yet $w^T\frac{1}{v}\geq 0$ and $w^TAA^Tv\leq \sqrt{w^TAA^Tw \times v^TAA^Tv}$, so the inequality becomes
$v^TAA^Tv  + t (v-w)^T(v-w)\leq M + \sqrt{w^TAA^Tw \times v^TAA^Tv}$

Now, we can assume $w^TAA^Tw=M$ so $v^TAA^Tv  + t (v-w)^T(v-w)\leq M + \sqrt{M} \sqrt{v^TAA^Tv}$. 

This implies, first, that $v^TAA^Tv \leq M + \sqrt{M} \sqrt{v^TAA^Tv}$. So, in particular $v^TAA^Tv\geq 4M$ is impossible because it will lead to $4M\leq M+2M$ which is false.

Then, $t (v-w)^T(v-w)\leq M +\sqrt{M v^TAA^Tv} - v^TAA^Tv\leq \max_u (M + \sqrt{M}u-u^2)\Rightarrow u = \frac{\sqrt{M}}{2}$.

So finally, $t (v-w)^T(v-w)\leq M +\frac{M}{2}-\frac{M}{4} = \frac{5}{4}M$.
So, $t \frac{(v-w)^T(v-w)}{2} \leq M$.

So the lemma 2 applies.
In addition, $\psi(w) =\psi(z(0)) = 0$.
Moving from $w$ which is close to $v^*(t(0))$ to $v^*(t(1))$, and from $v^*(t(1))$ to $v^*(t(2))$, and so on until $v^*(t(K))$ costs $O(K)$.

\subsubsection{Termination}
Obviously, the relevancy of starting a path following from $w$ depend on the value of $t(0)$ as the value of $K$ is $O(\sqrt{M}\log(\frac{t(0)}{t_f}))$.

As pointed in previous subsection $AA^Tv + t (v-w)=\frac{1}{v}$ when minimizing $G_w$ for a given $t$ in particular $AA^Tv\geq \frac{1}{v}-tv$.

But, $v^TAA^Tv\leq 4M$, so, $v\leq 2\sqrt{Mx^Tx}\mathbf{1}$.
In particular $AA^Tv\geq (\frac{1}{2\sqrt{Mx^Tx}}-t \times 2\sqrt{Mx^Tx})\mathbf{1}$, so as soon $t\leq \frac{1}{4M x^Tx}$, any minima for a given $t$ is a solution of the original problem $AA^Tv>\mathbf{0}$.

\subsubsection{Complexity}
Let $t(0)$ be the minimal $t$ for which $w$ is close to the minimum regarding $v$ of $G_w(v,t(0))$, and, let $K = O(\sqrt{M}\log(\frac{t(0)}{t_f}))$ with $t_f \leq \frac{1}{4Mx^Tx}$, then, one can recover a solution of $AA^Tv>\mathbf{0}$ problem by following the path created from $w$ using classical path following way in less than $O(K)$.

$t_f$ is somehow a constant of the problem, $t_0$ can not be bounded a priori but can be evaluated on the fly (distinguishing between relevant and irrelevant $w$).
So, of course, the idea is to alternate minimization of $F$ by direct Newton descent and by following the path induced by $G_w$ when finding a relevant $w$ (i.e. with a small $t(0)$).

\end{document}